\documentclass[twocolumn,showpacs,preprintnumbers,amsmath,amssymb]{revtex4-1}
\usepackage{epsfig}
\usepackage{graphicx}
\usepackage{epsfig}
\usepackage{color}
\usepackage{rotating}
\usepackage{amssymb}
\usepackage{amsmath}
\usepackage{graphicx}
\usepackage{tikz-cd} 
\usetikzlibrary{matrix,arrows,decorations.pathmorphing}

\def\beq{\begin{eqnarray}}
\def\eeq{\end{eqnarray}}

\newcommand{\be}{\begin{equation}}
\newcommand{\ee}{\end{equation}}
\newcommand{\bea}{\begin{eqnarray}}
\newcommand{\eea}{\end{eqnarray}}

\begin{document}

\title{A unified picture of strong coupling stochastic thermodynamics and time reversals}

\author{Erik Aurell}
\email{eaurell@kth.se}
\affiliation{
KTH -- Royal Institute of Technology,  AlbaNova University Center, SE-106 91~Stockholm, Sweden\\
Depts. Computer Science and Applied Physics, Aalto University, Espoo, FIN-00076 Aalto, Finland
}

\begin{abstract}
Strong-coupling statistical thermodynamics is formulated as Hamiltonian dynamics
of an observed system interacting with another unobserved system (a bath).
It is shown that the entropy production functional of stochastic thermodynamics,
defined as the log-ratio of forward and backward system path probabilities,
is in one-to-one relation with the log-ratios of joint initial conditions of the system and the bath.
A version of strong-coupling statistical thermodynamics
where the system-bath interaction vanishes at the beginning and the end
of a process is, as is also weak-coupling stochastic thermodynamics, related the bath 
initially in equilibrium by itself.
The heat is then the change of bath energy over the process, and 
it is discussed when this heat is a functional of system history alone.
The version of strong-coupling statistical thermodynamics introduced by Seifert and Jarzynski is related
to the bath initially in conditional equilibrium with respect to the system.
This leads to heat as another functional of system history
which needs to be determined by thermodynamic integration.
The log-ratio of forward and backward system path probabilities
in a stochastic process is finally related to log-ratios of initial conditions
of a combined system and bath. It is shown that the entropy production
formulas of stochastic processes under general class of time reversals
are given by differences of bath energies in a larger underlying Hamiltonian system. 
The paper highlights the centrality of time reversal
in stochastic thermodynamics, also in the case of strong coupling.
\end{abstract}


\maketitle

\section{Introduction}
\label{sec:Introduction}
Stochastic Thermodynamics 
describes mesoscopic systems
which can be controlled individually
while also interacting with a surrounding uncontrolled environment,
here for brevity called a bath.
Work done on such systems is as in classical macroscopic thermodynamics 
the total change in energy of the system and the bath
during a process.
In a general setting this could depend on the bath, but for
conservative dynamics where only the 
system Hamiltonian $H_S$ depends explicitly on time,
work defined this way reduces to $\int \partial_t H_S dt$,
a functional of the system history only~\cite{Jarzynski11-review,sekimoto-book,Searles08-review,Esposito09-review}. 

In its standard formulation Stochastic Thermodynamics 
assumes that the energy 
stored in the coupling between the system and the bath
is negligible compared to the system energy.
The internal energy change can then be taken to be the change of system energy only,
 and as work is then a quantity determined by system history only.
Heat can similarly be taken to be the change of bath energy.
By itself this is not measurable on the system, but it can be deduced from the
system history in many standard models in non-equilibrium physics,
in particular for Master Equations (for discrete states) and
for Langevin equations (for continuous states).
Work, heat and change in internal energy then obey a trajectory-wise First Law
where all three quantities are measurable functionals of the system history.
The theoretical and fundamental interest in Stochastic 
Thermodynamics stems to a considerable extent from work and heat
also satisfying exact equalities collectively known as
fluctuation relations~\cite{Jarzynski11-review,seifert12-review}.

``Strong coupling'' refers the setting where the variations of the energy stored in the coupling between the system and the bath
are comparable to or greater than the variations in system energy.
It is not obvious if such a change should be counted with the change of bath energy
as heat, or if it should be counted with the change of system energy
as an internal energy change, or if its variation should 
somehow be split between the two.
In the related quantum problem internal energy has in fact 
in different publications been assumed to
include none, half and all of the system-bath interaction energy,
for a recent critical discussion, see~\cite{PhysRevB.92.235440}.
It is therefore not obvious that there is a meaningful trajectory-wise First Law in
strong-coupling Statistical Thermodynamics, nor if there are
meaningful strong-coupling fluctuation relations.
The issue was first raised in~\cite{CohenMauzerall04}
and answered for Jarzynski Equality (JE) 
soon after in~\cite{Jarzynski2004}, where this fluctuation relation was restated as
\begin{equation}
\label{eq:JE-strong}
\left< e^{-\beta\delta W}\right>_{eq} = e^{-\beta \Delta \tilde{F}_S}.
\end{equation}
In above $\beta=\frac{1}{k_B T}$ is the inverse temperature, $\delta W$ is the work, and the average 
is over realizations starting from a joint equilibrium of the system and the bath.
The left hand side is hence the same as in standard Stochastic Thermodynamics
and measurable on the system alone.
The quantity $\tilde{F}_S$ on the right hand side is on the other hand
a free energy at mean force~\cite{Onsager1933,Kirkwood1935,Ford1988,Hanggi2008,Jarzynski2004,PhysRevE.79.051121,Campisi11-review}.
This depends on the equilibrium
state of the system and the bath together. It is not
measurable on the system alone, but has to be deduced
by thermodynamic integration \textit{i.e.} by following changes in $\tilde{F}_S$ 
as temperature or other parameters are varied.
Importantly the right hand side of (\ref{eq:JE-strong}) however does not
depend on the protocol for changing the system energy $H_S$
while the left-hand side does.
This shows that there is a meaningful strong-coupling JE, and also that fluctuating
strong-coupling work is a meaningful quantity.

Other strong-coupling fluctuation relations have been slower to obtain.
In fact, up to the recent proposals in~\cite{Seifert2016} and \cite{Jarzynski2017}
there was no strong-coupling definition of 
total entropy change in the combined system and bath that would satisfy 
the Integral Fluctuation Theorem (IFT)
\begin{equation}
\label{eq:IFT-strong}
\left< e^{-\Delta S_{TOT}}\right> = 1.
\end{equation}
Heat would be related to such a quantity as
$\delta Q=\beta\left(\Delta S_{TOT}-\Delta S_{S}\right)$
where a general definition of $\Delta S_{S}$, the entropy change of the
system, has also been lacking.
The proposal of~\cite{Seifert2016}, to be discussed below, 
was criticized in~\cite{TalknerHanggi2016}, where the authors reached the conclusion that 
open system trajectories only specify work and not heat.
Following upon \cite{Seifert2016,Jarzynski2017}
two important steps were later taken in~\cite{StrasbergEsposito2017}
where the proposal was derived by a time-scale separation argument (coarse-graining),
and in~\cite{MillerAnders2017}, where it was related to a time reversal.

The first goal of this paper is to re-state the issue of strong-coupling
thermodynamics as one of time reversals in a combined system and bath.
It will emerge that the entropy production functional 
of stochastic thermodynamics
is equal to the log-ratio of probabilities of initial states in the larger system.
Although quite simple this result was to the author's knowledge 
first explicitly stated quite recently~\cite{MillerAnders2017}.

Entropy production functionals as log-ratios
lead to fluctuation relations as ``tautologies''~\cite{Maes99,Gawedzki13}.
The second goal of this paper is hence to show that different 
initial probabilities and different
time reversals of a system and a bath lead to different 
entropy production functionals which all satisfy 
fluctuation relations. 
This also gives a new perspective on entropy production and time
reversals in stochastic differential systems, whenever these can be
seen as the effective dynamics of a system also interacting with a bath.

The paper is organized as follows. In Section~\ref{sec:forward-backward} I
relate ratios of forward and backward path probabilities of the system to 
initial probability distributions of the combined system and bath
in the forward and backward process. In Section~\ref{sec:examples}
I discuss three different examples. 
In the first system-bath interaction is assumed to vanish at 
the beginning and the end of the process,
and the bath is initially in equilibrium, while the system state can be arbitrary. 
This gives an additional term in the work, as recently discussed 
at length in~\cite{Aurell2017}, but heat is simply the change of bath energy,
the same as in weak-coupling statistical thermodynamics.
The second example is a re-formulation of
the approach of~\cite{Seifert2016,Jarzynski2017,StrasbergEsposito2017,MillerAnders2017}
where the bath is initially in conditional equilibrium with respect to the system.
The last example is finally, as in the discussion
around (\ref{eq:JE-strong}) above, of the case when the system and the bath are assumed initially
jointly in equilibrium. 
This leads to formulas for heat which at first glance look
unfamiliar, but which can be reduced to the case of conditional equilibria.
In Section~\ref{sec:stochastic}
I consider entropy production and time general reversals in stochastic dynamics
when that dynamics results from interaction with a bath,
and show that related entropy production functions equal the differences of bath energies in units of $k_B T$.
In Section~\ref{sec:Conclusions} I discuss and compare the results.
Three appendices contain technical details or material which is either standard or already
presented elsewhere.

\section{Forward-backward path probabilities and baths}
\label{sec:forward-backward}
I will assume that the system and the bath together are one big closed conservative system.
The total Hamiltonian is hence
\begin{equation}
\label{eq:H}
H_{TOT}(x,y) = H_S(x) + H_I(x,y) + H_B(y)
\end{equation}
where the three parts refer to the system, the interaction and the bath, respectively.
The phase space of the system is parametrized by $x$ (coordinates and momenta of the system) and the phase space
of the bath is parametrized by $y$ (coordinates and momenta of the bath).
I will assume either that only $H_S$ depends explicitly on time,
or that only $H_S$ and $H_I$ depend explicitly on time. 
The initial state of the system and the bath is $\rho_i(x^i,y^i)$
where the subscript indicates ``initial''.
Special classes will be considered later.

Let us assume that the system has $D$ degrees of freedom and the bath $N$ degrees of
freedom. Observing the system at $n=\frac{N}{D}$ time points $t_1,t_2,\ldots,t_n$ should
generically give the same information as observing the bath at the initial time $t_i$.
We may therefore postulate an equivalence between the probability distribution $\rho_i$
over initial conditions of the total system, and the joint probability distribution
$P^F(x_0,x_1,\ldots,x_n)$ of the coordinates and momenta of the system at time points
$t_i=t_0,t_1,t_2,\ldots,t_n=t_f$. By the law of conservation of probability this equivalence is
\begin{equation}
\label{eq:P-F-1}
P^F(x_0,x_1,\ldots,x_n) \prod_{k=0}^n dx_k =  \rho_i(x^i,y^i) dx^i dy^i
\end{equation} 
The shift from $x^i,y^i$ to $x_0,x_1,\ldots,x_n$ is a change of variables.
Eq.~(\ref{eq:P-F-1}) can therefore also be written
\begin{equation}
\label{eq:P-F-2}
P^F(x_0,x_1,\ldots,x_n)  =  \rho_i(x^i,y^i) |\frac{\partial (\{x_k\}_{k=0}^n)}{\partial (x^i,y^i)} |^{-1}
\end{equation} 
where the second term is Jacobian of the transformation.

Let us now consider a time-reversed process parametrized by a reversed time $t^*=t_f-t$.
This process starts at $t^*_i =0$ ($t=t_f$) and runs until  $t^*_f=t_f-t_i$ ($t=t_i$).
The general concept of time reversal in stochastic thermodynamics was discussed
in great detail by Ch\'etrite and Gaw\c{e}dzki in~\cite{ChG07}.
I will assume that time reversal is implemented by a functional ${\cal I}$ 
such that the time-reversed coordinates $(x^*_{t*},y^*_{t*})$ 
are ${\cal I}(x_{t},y_{t})$ and the time-reversed Hamiltonian $H^*_{t^*}$ is ${\cal I} H_t$.
Time-reversed dynamics is thus Hamiltonian dynamics in the coordinates
$(t^*,x^*_{t*},y^*_{t*})$ with Hamiltonian $H^*_{t^*}(x^*_{t*},y^*_{t*})$.  
The time-reversal functional is assumed to have the following general properties:
\begin{description}
\item[Involution] ${\cal I}$ is an involution on $(x,y,H)$, \textit{i.e} $({\cal I})^2=\mathbf{1}$.
\item[Separability] ${\cal I}$ acts separately on the system \textit{i.e} $\left[{\cal I}(x,y)\right]_{system}= {\cal I}x$. 
\item[Volume preservation] ${\cal I}$ separately preserves system phase space volume and bath phase space volume.
\end{description}
A main example which satisfies all above is standard
time inversion of all the generalized coordinates as $q^*_{t*}=q_t$
and the generalized momenta as $p^*_{t*}=-p_t$, and
Hamiltonian, when there is no magnetic field, as $H^*_{t*}=H_t$. 
When there is a non-zero magnetic field
time reversal can be done by changing the sign of magnetic field~\cite{Kubo1957},
but other time reversals are also possible~\cite{Bonella2017}.

Let the initial density of the time-reversed process be 
$\rho_i^*$. Then the backward path probability satisfies
\begin{equation}
\label{eq:P-B}
P^B(x^*_0,\ldots,x^*_n) \prod_{k=0}^n dx^*_k =  \rho_i^*((x^*)^i,(y^*)^i) d(x^*)^i d(y^*)^i
\end{equation}
Combining (\ref{eq:P-F-1}) and (\ref{eq:P-B}) one has
\begin{equation}
\label{eq:P-ratio-general}
\frac{P^F(x_0,\ldots,x_n)}{P^B(x^*_0,\ldots,x^*_n)}
= \frac{\rho_i(x^i,y^i)}{\rho_i^*((x^*)^i,(y^*)^i)}
\left|\frac{\frac{\partial (\{x^*_k\}_{k=0}^n)}{\partial ((x^*)^i,(y^*)^i)}} {\frac{ \partial (\{x_k\}_{k=0}^n)}{\partial (x^i,y^i) }} \right|
\end{equation}
The ratio of Jacobians in (\ref{eq:P-ratio-general})
can be combined with 
$\frac{\partial (\{x^*_k\})}{\partial (\{x_k\})}$ and
$\frac{\partial ((x^*)^i,(y^*)^i)}{\partial (x^f,y^f)}$
both of which have absolute value one.
The absolute ratio of in Jacobians (\ref{eq:P-ratio-general})
therefore has the same value 
as 
$|\frac{\partial (x^f,y^f)}{(x^i,y^i)}|^{-1}$, 
which is one, because Hamiltonian dynamics preserves total phase space volume

Instead of (\ref{eq:P-ratio-general}) we therefore have much more simply
\begin{equation}
\label{eq:P-ratio-special}
\frac{P^F(x_0,x_1,\ldots,x_n)}{P^B(x^*_0,x^*_1,\ldots,x^*_n)}
= \frac{\rho_i(x^i,y^i)}{\rho_i^*((x^*)^i,(y^*)^i) }
\end{equation}
Equation (\ref{eq:P-ratio-special}) says that for 
classical conservative systems path probabilities are only 
consequences of uncertainties in the initial conditions,
and ratios of path probabilities are
given by ratios of probabilities of initial 
conditions.

\section{Scenarios for strong-coupling heat}
\label{sec:examples} 
In this Section I will give self-contained descriptions
of three scenarios.
The scenarios differ only in what is assumed
for the initial states $\rho_i(x^i,y^i)$
and $\rho_i((x^*)^i,(y^*)^i)$.
The descriptions end with a summary of what strong-coupling heat
has to be in each scenario to satisfy the 
Integrated Fluctuation Theorem (\ref{eq:IFT-strong}).

\subsection{Factorized equilibria with time-dependent system-bath coupling}
\label{sec:factorized} 
In standard Stochastic Thermodynamics the interaction between the system and the bath
is weak and the bath is initially in equilibrium by itself.
The smallest deviation from this scenario that allows to treat also
strong coupling is to assume that the interaction is time-dependent,
and vanishing at the beginning and the end of the process.
As then both $H_S$ and $H_I$ depend explicitly on time
the work is
\begin{eqnarray}
\Delta H_{TOT} &=& \delta W^{(J)}+ \delta W_{if} \nonumber \\
              &=&  \int \partial_t H_S dt + \int \partial_t H_I dt
\end{eqnarray}
The first term in above is as in (\ref{eq:JE-strong}) the Jarzynski work
while the second term was introduced in~\cite{Aurell2017}.
It is a functional of system history only for
some models of the bath and the system-bath interaction. In particular 
it is however so for the Zwanzig model (Caldeira-Leggett model)
which leads to Kramers-Langevin system dynamics~\cite{Zwanzig,CaldeiraLeggett83a,Grabert88}.
A summary with some extensions is given in Appendix~\ref{app:boundary-work}.

The factorized initial conditions, where the bath is in equilibrium,
are 
\begin{equation}
\label{eq:def-conditional-2}
\rho_i(x^i,y^i) = \rho_S^i(x^i)\rho_B^{eq}(y^i) 
\end{equation}
where the system state $\rho_S^i(x^i)$ can be anything and 
\begin{equation}
\label{eq:bath-eq-def}
\rho_B^{eq}(y) = e^{-\beta\left(H_B(y)-F_B\right)}. 
\end{equation}
There is no dependence on the interaction Hamiltonian in (\ref{eq:bath-eq-def}) 
since that has been assumed to vanish at the beginning of the process.

The initial conditions of the backwards process 
are analogously
\begin{equation}
\label{eq:def-conditional-2}
\rho_i((x^*)^i,(y^*)^i) = \rho_S^{i,*}((x^*)^i)  \rho_B^{eq}((y^*)^i) 
\end{equation}
which gives an entropy production
\begin{eqnarray}
\Delta S_{TOT}^{(fact.eq.)} &=& \log \frac{P^F(x_0,x_1,\ldots,x_n)}{P^B(x^*_0,x^*_1,\ldots,x^*_n)} \nonumber \\
&=& \log \rho_S^i - \log\rho_S^{i,*}  \nonumber \\
&& + \log \rho_B^{eq}(y^i)  - \log \rho_B^{eq}((y^*)^i) 
\label{eq:entropy-1-2}
\end{eqnarray}
In Section~\ref{sec:stochastic} and Appendix~\ref{app:stochastic-details}
I consider a class of examples where the comparison is
made between $\log \rho_B^{eq}((y^*)^i)$ and $\log \rho_B^{eq}(y^i)$
and where $(y^*)^i$ is determined from the whole system path.
In a general setting $(y^*)^i$ will hence not be a simple transformation
of $y^f$ only. Assuming here that 
the equilibrium state of the bath is time-reversal invariant, that is $\rho_B^{eq}((y^*)^i)=\rho_B^{eq}(y^f)$,
which holds for the ``canonical'' time reversal of Section~\ref{sec:stochastic},
the difference in the last line in (\ref{eq:entropy-1-2}) is 
$\beta\Delta H_B$, the change in bath energy in units of $k_B T$.

If further the initial state of the time-reversed system ($\rho_S^{i,*}$) is
identical to the final state of of the system going forwards ($\rho_S^f$)
one recognizes in (\ref{eq:entropy-1-2}) from standard Stochastic Thermodynamics
the stochastic entropy $-\Delta \log\rho$, 
the negative log-change in probability density from an initial position at the initial time to a final position~\cite{seifert12-review}.
It is simple to then re-write (\ref{eq:entropy-1-2}) as
\begin{eqnarray}
\label{eq:entropy-2-2}
\Delta S_{TOT}^{(fact.eq.)} &=& -\Delta\log P +\beta\left(\delta W^{(J)}+ \delta W_{if}\right) \nonumber \\
                  && - \beta\Delta H_S
\end{eqnarray}
The heat functional in this scenario is thus
\begin{eqnarray}
\label{eq:heat-2-3}
\delta Q^{(fact.eq.)} &=& \delta W^{(J)} + \delta W_{if} - \Delta H_S = \Delta H_B
\end{eqnarray}
Since interaction energy has been assumed to vanish at the boundaries,
heat is only the change in bath energy during the process,
the same as in standard (weak coupling) stochastic thermodynamics.
If $\delta Q^{(fact.eq.)}$ in (\ref{eq:heat-2-3}) is a functional measurable on the system alone
however depends on the second term $\delta W_{if}$, see Appendix~\ref{app:boundary-work}.

\subsection{Conditional equilibria with time-reversal symmetric states}
\label{sec:conditional} 
Next I turn to the approach
of~\cite{Seifert2016},~\cite{Jarzynski2017} and~\cite{MillerAnders2017}.
Only $H_S$ now depends explicitly on time, and the work functional is
as in (\ref{eq:JE-strong}) only the Jarzynski work
\begin{equation}
\label{eq:Jarzynski}
\Delta H_{TOT} = \delta W^{(J)}=
\int \partial_t H_S dt
\end{equation}
The bath is assumed to be initially in equilibrium conditional
of the system:
\begin{equation}
\label{eq:def-conditional}
\rho_i(x^i,y^i) = \rho_S^i(x^i)\sigma(y^i | x^i) 
\end{equation}
where $\rho_S^i(x^i)$ can be anything and
\begin{equation}
\label{eq:def-sigma}
\sigma(y^i | x^i) = \frac{e^{-\beta\left(H_I(x^i,y^i)+ H_B(y^i)\right)}}
                        {\int dy' e^{-\beta\left(H_I(x^i,y')+ H_B(y')\right)}}
\end{equation}
The initial conditions of the time-reversed
process are also such that the bath is 
in equilibrium conditional
to the system, and adopting 
analogous assumptions to above (also stated in~\cite{MillerAnders2017}) 
I will assume that the conditional distribution of the bath is time-reversal symmetric. 
This means
\begin{equation}
\label{eq:def-conditional}
\rho_i((x^*)^i,(y^*)^i) = \rho_S^{i,*}((x^*)^i) \sigma(y^f | x^f) 
\end{equation}
with the same conditional probability as in (\ref{eq:def-sigma}).
The total entropy change is then
\begin{eqnarray}
\Delta S_{TOT}^{(cond.eq.)} &=& \log \frac{P^F(x_0,x_1,\ldots,x_n)}{P^B(x^*_0,x^*_1,\ldots,x^*_n)} \nonumber \\
&=& \log \rho_S^i - \log \rho_S^{i,*} \nonumber \\
&& + \log \sigma(y^i | x^i) - \log \sigma(y^f | x^f) 
\label{eq:entropy-1}
\end{eqnarray}
In the same setting as in the previous section where
initial state of the system going backwards ($\rho_S^{i,*}$) is
the same as final state of the system going forwards ($\rho_S^f$)
it was shown in \cite{Seifert2016} that
(\ref{eq:entropy-1}) can be re-written as
\begin{eqnarray}
\label{eq:entropy-2}
\Delta S_{TOT}^{(cond.eq.)} &=& \Delta\tilde{s}_S + \beta\delta W^{(J)} - \beta \Delta \tilde{u}_S
\end{eqnarray}
where $\tilde{u}_S$ is an energy-like function,
 $\tilde{f}_S$ is the constant in a Gibbs-Boltzmann-like
distribution $P^{(cond.eq.)}=e^{\beta\left(\tilde{f}_S-\tilde{u}_S\right)}$
and $\tilde{s}_S=-\log P^{(cond.eq.)}$ is the corresponding entropy-like quantity. 
For completeness this derivation is repeated in Appendix~\ref{app:Seifert-Anders}.

The heat functional in this scenario is thus
\begin{eqnarray}
\label{eq:heat-2-3}
\delta Q^{(cond.eq.)} &=& \delta W^{(J)} - \Delta \tilde{u}_S
\end{eqnarray}
As $\tilde{F}_S$ in (\ref{eq:JE-strong})
the quantities $\tilde{f}_S$, $\tilde{u}_S$ and
$\tilde{s}_S$ depend on the bath. 
Parameter variation, 
\textit{i.e.} thermodynamic integration,
is needed to determine an arbitrary constant in
$\tilde{u}_S$ and $\tilde{f}_S$ 
which would otherwise render
(\ref{eq:entropy-2})
and (\ref{eq:heat-2-3}) indeterminate.

The explicit form of $\tilde{u}_S$, re-derived in Appendix~\ref{app:Seifert-Anders} and stated in 
(\ref{eq:U-Seifert}), is $H_S -\partial_{\beta} \log\left<e^{-\beta H_I}\right>_B$
where $\left<\cdots\right>_B$ indicates average with respect to the a Gibbs state $e^{\beta (F_B- H_B)}$.
The change $\Delta \tilde{u}_S$ hence includes the change
in system energy $\Delta H_S$ and the change in average both bath and interaction energy
with respect to a conditional bath Gibbs state $e^{\beta (F'_B- H_B-H_I)}$
($H_I$ and $F'_B$ depend on system coordinate).
The heat in (\ref{eq:heat-2-3}) includes the corresponding 
fluctuating quantities. It is quite interesting that 
the proposal in~\cite{Seifert2016} hence does not reduce to any of
the simpler earlier suggestions that counted in the heat 
some definite fractions of 
respectively $\Delta H_B$ and $\Delta H_I$. 

\subsection{Joint equilibrium of the system and the bath}
\label{sec:total} 
The last scenario 
adheres closely to the
the equilibrium strong-coupling theory and several earlier
contributions~\cite{Onsager1933,Kirkwood1935,Ford1988,Hanggi2008,Jarzynski2004,PhysRevE.79.051121,Campisi11-review}.
Of the three terms in (\ref{eq:H}) again only the system Hamiltonian
$H_S$ depends explicitly on time and the work is given by (\ref{eq:Jarzynski}).
The assumption is now that the bath and the system are jointly in equilibrium at the
beginning of the process
\begin{equation}
\label{eq:def-conditional-3}
\rho_i(x^i,y^i) = \rho_{TOT}^{eq}(x^i,y^i) =\frac{1}{Z_{TOT}^i}e^{-\beta H^i_{TOT}}   
\end{equation}
The initial conditions of the backwards process 
are analogously taken to be
\begin{equation}
\label{eq:def-conditional-backwards-3}
\rho_i((x^*)^i,(y^*)^i) = \frac{1}{Z_{TOT}^f}e^{-\beta H^f_{TOT}}    
\end{equation}
From (\ref{eq:P-ratio-special}) we then have 
\begin{eqnarray}
\Delta S_{TOT}^{(tot.eq.)} &=& \log \frac{P^F(x_0,x_1,\ldots,x_n)}{P^B(x^*_0,x^*_1,\ldots,x^*_n)} \nonumber \\
\label{eq:entropy-1-3}
&=& \beta \delta W^{(J)} +\log Z_{TOT}^f - \log Z_{TOT}^i
\end{eqnarray}
The Jarzynski work is a functional of the system history and gives, for this
scenario, all the coordinate dependence. The statistical properties of 
$\Delta S_{TOT}^{(tot.eq.)}$ and $\delta W^{(J)}$ are therefore in this scenario the same.

The last two terms (constants) in (\ref{eq:entropy-1-3}) 
can be referred to the total free energy with respect to that of the bath alone 
\begin{equation}
\label{eq:F-Hanggi}
\tilde{F}_S = \frac{1}{\beta}\log\frac{Z_B}{Z_{TOT}}=F_{TOT}-F_B
\end{equation}
and are thus the change of a free energy at mean force, as already used 
in (\ref{eq:JE-strong}) above:
\begin{equation}
\log Z_{TOT}^f - \log Z_{TOT}^i = -\beta\Delta \tilde{F}_S
\end{equation}

The free energy at mean force can be written
\begin{equation}
\tilde{F}_S = \tilde{U}_S -\frac{1}{\beta} \tilde{S}_S
\end{equation}
where the internal energy (or potential) at mean force is 
\begin{eqnarray}
\tilde{U}_S &=& \partial_{\beta}\left(\beta \tilde{F}_S\right) = U_{TOT}-U_B 
\end{eqnarray}
and the corresponding entropy is
\begin{eqnarray}
\tilde{S}_S &=& \beta\left(\tilde{U}_S-\tilde{F}_S\right) = S_{TOT}-S_B 
\end{eqnarray}

With these conventions (\ref{eq:entropy-1-3})
can be re-written
\begin{eqnarray}
\label{eq:entropy-2-3}
\Delta S_{TOT}^{(tot.eq.)} &=& \Delta\tilde{S}_S + \beta\delta W^{(J)} - \beta \Delta \tilde{U}_S
\end{eqnarray}
and the heat functional is
\begin{eqnarray}
\label{eq:heat-3}
\delta Q^{(tot.eq.)} &=& \delta W^{(J)} - \Delta \tilde{U}_S
\end{eqnarray}

To compare (\ref{eq:heat-3}) to (\ref{eq:heat-2-3}) we must recognize that the time-reversals are
qualitatively different. The heat in (\ref{eq:heat-2-3}) was derived under the assumption that
the initial state of the system in backward process is the same as final state of the forward process.
This is not the same as in (\ref{eq:heat-3}) where the 
initial state of the system in the backward process is 
the marginal of a total equilibrium state,
while the final state
of the forward process is generally something else.
To compare we must instead go back to the total entropy productions in
(\ref{eq:entropy-1}) and (\ref{eq:entropy-1-3})
and identify the initial system states of the forward and backward states
in (\ref{eq:entropy-1}) as $=e^{\beta\left({\cal F}_S -{\cal H}_S\right)}$,
where ${\cal H}_S$ and ${\cal F}_S$ are the potential and free energy at at mean force of Onsager and 
Kirkwood~\cite{Onsager1933,Kirkwood1935,Jarzynski2004,PhysRevE.79.051121,Campisi11-review}.
With this (\ref{eq:entropy-1}) reduces to (\ref{eq:entropy-1-3})
\begin{eqnarray}
&&\Delta S_{TOT}^{(cond.eq.-red.)} = \beta \Delta {\cal H}_S - \beta \Delta {\cal F}_S = \nonumber \\
&&\quad = \beta \Delta H_{TOT} - \beta \Delta H_{S} +\Delta \log\left<e^{-\beta H_I}\right>_B   
\label{eq:entropy-1-reduced}
\end{eqnarray}
where 
in the second line I have used (\ref{eq:log-terms}) from Appendix~\ref{app:Seifert-Anders}.

\section{Time reversals in stochastic dynamics}
\label{sec:stochastic}
In this Section the focus is not on strong coupling.
The interaction will hence be taken weak, or assumed to depend on time as
in Section~\ref{sec:factorized}.
The focus is instead on using the 
general result in Section~\ref{sec:forward-backward}
to give a new perspective on time reversals in stochastic dynamics~\cite{ChG07}.
To lighten the presentation technical details are given in Appendix~\ref{app:stochastic-details}.

It is well known that a 
Kramers-Langevin equation 
$\dot{x}=\frac{p}{m}$ and $\dot{p}=-\partial_x V(x,t) - \gamma\frac{p}{m} + \sqrt{2k_B T\gamma}\xi$
can be derived from the total Hamiltonian dynamics of a system interacting
linearly with a bath of harmonic oscillators
which are initially in thermal equilibrium~\cite{Zwanzig,Bez1980,Hanggi1997}.
\textit{Complete time reversal} refers to standard
time inversion of all the coordinates and momenta, of
both the bath and the system.
On the level of the system this is a process conditioned by the future,
that at the final time the bath will be in equilibrium, and is therefore 
not a Markov process.
It follows immediately from (\ref{eq:P-ratio-special}) that the entropy production in such
a time reversal is zero
because the right-hand side of (\ref{eq:P-ratio-special})
can also be written $\frac{\rho_f(x^f,y^f)}{\rho_i^*((x^*)^i,(y^*)^i)}$ (preservation of phase space volume)
and this ratio is one (time reversal preserves phase space volume).
This is logical because when the motion of both the system and the bath are reversed they will evolve back to their
initial state, and no information will be lost.

The closest to complete time reversal 
defined on the level of the system is 
\textit{natural time reversal}~\cite{ChG07}.
This is standard time reversal on the system
and transforming the dynamics to
$\frac{dx^*}{dt^*}=\frac{p^*}{m}$ and $\frac{dp^*}{dt^*}=-\partial_x V(x,t_f-t) + \gamma\frac{p^*}{m} + \sqrt{2k_B T\gamma}\xi^*$,
where $\xi^*$ is a noise with the same statistical properties as $\xi$.
The anti-friction ($\gamma\frac{p^*}{m}$) shows that this equation does not originate 
from the system interacting with a bath initially in thermal equilibrium.
In the other direction it was shown in~\cite{ChG07} that the entropy production 
associated to natural time reversal is $(t_f-t_i)\gamma /m$; natural time reversal is therefore different from complete time reversal.
For completeness a sketch of a derivation of this fact is given Appendix~\ref{app:stochastic-details}.

We turn now instead to \textit{canonical time reversal}~\cite{ChG07}, where
the backward process also obeys a Kramers-Langevin with positive friction:
$\frac{dx^*}{dt^*}=\frac{p^*}{m}$ and $\frac{dp^*}{dt^*}=-\partial_x V(x,t_f-t) - \gamma\frac{p^* }{m} + \sqrt{2k_B T\gamma}\xi^{'}$,
and $\xi'$ again is a noise with the same statistical properties as $\xi$.
It is convenient to consider a wider class of \textit{general time reversals},
introduced in \cite{ChG07}
by splitting the drift field (time derivative of system coordinate).
We split the system potential
in two parts that transform differently, $V_t=V_t^+ + V_t^-$
and the time-reversed total Hamiltonian $H^*_{t^*}$
will contain the piece $V_{t^*}^{*,+} - V_{t^*}^{*,-}=V_{t}-2 V_{t}^-$.
Canonical time reversal is the case when $V^-=0$.
The system equation under such general time reversal is
$\frac{dx^*}{dt^*}=\frac{p^*}{m}$ and $\frac{dp^*}{dt^*}= - \partial_x V + 2\partial_x V^{-} - \gamma\frac{p^*}{m} + \sqrt{2k_B T\gamma}\xi^{''}$.
Introducing the notation of \cite{ChG07} that $u_+= -\gamma p/m -\partial_x V^-$
is the part of the drift field that transforms as a vector
and $u_-= -\partial_x V^+$ is the part that transforms as a pseudo-vector,
and identifying $D=k_B T\gamma$ as the diffusion coefficient, one has
\begin{eqnarray}
\Delta S_{TOT} &=& \log\frac{P^F}{P^B} \nonumber \\
&&\quad = -\log\Delta P + \int \left(\dot{p}-u_-\right)\frac{1}{D} u_+ dt
\label{eq:CG-formula}
\end{eqnarray}
which is a main result of \cite{ChG07}, adapted to this situation.
Using explicit expressions for the dynamics of the continuum of harmonic oscillators
that make up the bath it is on the other hand straight-forward to show that
\begin{equation}
H_B^*((y^*)^i) - H_B(y^i) 
= \int \left(\dot{p}-u_-\right)\frac{1}{\gamma} u_+ dt
\label{eq:CG-formula-heat}
\end{equation}
with the same definitions of $u_-$ and $u_+$ as above.
Detailed derivations of (\ref{eq:CG-formula}) and (\ref{eq:CG-formula-heat})  
are given in Appendix~\ref{app:stochastic-details}.
The entropy production formula under general time reversal 
is thus, in fact, the energy difference in a microscopic bath model in units of $k_B T$.
For canonical time reversal (\ref{eq:CG-formula-heat}) simplifies to
\begin{equation}
\hbox{Eq.~\ref{eq:CG-formula-heat} (canonical reversal) } 
= \int -\frac{p}{m}dp - \partial_x V dx
\label{eq:CG-formula-heat-canonical}
\end{equation}
where the right hand side equals the work ($\delta W = \Delta H_{TOT}$) minus the total change of
system energy ($\Delta H_{S}$). For this time reversal $H_B^*((y^*)^i)- H_B(y^i)$ hence equals
$H_B(y^f)- H_B(y^i)$, the change in bath energy in the forward process,
and $e^{-\beta H_B^*((y^*)^i)}=e^{-\beta H_B(y^f)}$.

The above examples extend naturally to when the
system-bath coupling is nonlinear in the system.
As already found in~\cite{Zwanzig} this leads to
friction term nonlinear in the system coordinate
and a noise term which satisfies an Einstein relation 
with the friction term. 
More recently perturbative solutions have 
been found when the bath is weakly an-harmonic~\cite{BhadraBanerjee2016,KrugerMaes2017}.
Although these contributions establish a form of fluctuation-dissipation theorems,
they can also be interpreted as showing that naive versions of
fluctuation-dissipation theorems do not hold. Hence, at least some general
diffusion processes where the noise terms do not satisfy an Einstein relation 
with the friction term also have representations in terms of explicit baths.

Time-reversals in over-damped stochastic systems, where 
the diffusion tensor $D$ can depend on the coordinate
effected by the noise ($dx=\cdots +\sqrt{2 D} dW$, $D=k_B T/\gamma$), 
can be embedded in the under-damped case discussed 
above ($dx=\frac{p}{m}dt$, $dp= \cdots -\gamma dx + \sqrt{2k_B T \gamma} dW$).
When temperature is constant
the over-damped limit gives no new contributions to the entropy
production~\cite{Celani12}.
Entropy production under a general time-reversal 
of an over-damped stochastic with possibly space-dependent friction coefficient $\gamma$
can hence also be related to energy change in a bath, as above. 

\section{Conclusions}
\label{sec:Conclusions}
Entropy production is related to irreversibility.
I have here considered the log-ratio of
forward and backward path probabilities of a system,
and shown them to be related to log-ratios of the initial state of
the total system (system and bath) in the forward 
and backward process.

Depending on what is assumed for the initial state of the bath,
one gets different entropy productions. This is 
not surprising because different initial states of the bath correspond to different
levels of control, and time reversal then leads to different loss
of information. 
Here I compare (\ref{eq:entropy-2}) and (\ref{eq:entropy-2-2}).
In both cases the initial state of the system can formally be anything.
In practice it is however reasonable to assume in the first case
either that the system and the bath are jointly in equilibrium
(discussed above in Section~\ref{sec:total}), or
that the system has been fixed for some time in position $x^i$
so that the bath will have had time to relax 
to conditional equilibrium $\sigma(y^i | x^i)$.
I hence assume that this is the scenario for both the forward and backward process.
Using the explicit expression of $\tilde{u}_S$
from (\ref{eq:U-Seifert}) we then have
\begin{eqnarray}
&&\delta Q^{(fact.eq.)} - \delta Q^{(cond.eq.)} =  \nonumber \\
&&\quad \delta W_{if} - \left< H_I + H_B\right>_{x^f} + \left< H_I + H_B\right>_{x^i} 
\label{eq:heat-diff-1-3-v2}
\end{eqnarray}
The difference in heat 
is hence in one part 
the extra work $\delta W_{if}$ needed 
change the system-bath interaction,
and in the other part the change in expected 
value of the bath energy and interaction energy,
conditioned on system state. For factorized equilibrium
this second term vanishes while for conditional bath equilibrium
it is counted in the change of internal energy. 
The two different forms 
of strong-coupling heat are hence mutually compatible.
The critique of~\cite{TalknerHanggi2016}
that strong-coupling heat is not a uniquely
defined concept can therefore partly be re-formulated as saying
that that different versions correspond to different physical situations.

Finally, I have in this work shown that the entropy production functional
of stochastic thermodynamics applied to diffusive systems 
defined as the log-ratio of path probabilities can be interpreted as the change of bath energy
in an underlying more detailed microscopic model.
This is a new connection between the mathematical and physical
notions of entropy production, and a further
strong argument in favor of the physical soundness
of stochastic thermodynamics. 

\section*{Acknowledgments}
I thank Rapha\"el Ch\'etrite for important clarifying remarks on time-reversals in stochastic dynamics
and Janet Anders, Peter H\"anggi and Udo Seifert for discussions and valuable comments.
This research was supported by the Academy of Finland through its Center of Excellence COIN,
grant no. 251170.

\appendix

\section{Work with time-dependent system-bath interaction}
\label{app:boundary-work}
This appendix summarizes the discussion in~\cite{Aurell2017} 
of time-dependent strong coupling, with some extensions.

I will now write the system as $x=(Q,P)$ and the bath as  $y=(q,p)$
and I will assume that the system and the bath only interact through
their generalized coordinates
\begin{equation}
H_{TOT} = H_S(Q,P,t) + H_I(Q,q,t) + H_B(q,p)
\end{equation}
where the explicit time dependencies have been indicated.
The equation of motion of the system is
\begin{equation}
\dot{Q}=\partial_P H_S(Q,P,t) \quad\hbox{(typically $=\frac{P}{M}$)}
\end{equation}
and 
\begin{equation}
\dot{P}=-\partial_Q H_S(Q,P,t) - \partial_Q H_I(Q,q,t)
\end{equation}
The second term, which depends on bath coordinate $q$, is a force acting on the system,
conventionally said to be from the bath on the system. For the Zwanzig (Caldeira-Leggett) model the bath is 
collection of harmonic oscillators and the interaction term is
\begin{equation}
H_I = -Qq C_q(t) + \frac{1}{2m_q \omega_q^2}Q^2 C_q^2(t)\quad\hbox{(Zwanzig)}
\end{equation}
In above $C_q(t)$ is the time-dependent interaction coefficient between the system and
bath oscillator $q$, $m_q$ and $\omega_q$ are the mass and angular frequency of that
oscillator, and the last term (which does not depend on $q$) is the Caldeira-Leggett counter-term.
The force from the bath on the system is then
\begin{equation}
-\partial_Q H_I = q C_q(t) - \frac{1}{m_q \omega_q^2}Q C_q^2(t) \quad\hbox{(Zwanzig)}
\end{equation}
It is well known that for an Ohmic bath with all $C_q$ constant this tends 
to the sum of the friction force and the random force in a Kramers-Langevin equation.
In~\cite{Aurell2017} was considered the situation where for all interaction coefficients
$C_q\propto \sqrt{\eta(t)}$ where $\eta(t)$ is a time-dependent friction coefficient.
In that setting the force from the bath on the system is
\begin{equation}
-\partial_Q H_I \approx -\eta \dot{Q} -\frac{\dot{\eta}}{2\eta}Q + \sqrt{\frac{2\eta}{\beta}}\xi
\quad\hbox{(from \cite{Aurell2017})}
\end{equation}
where $\xi$ is standard white noise.

From the structure of the interaction term it is now easy to determine
the second contribution to the work for the Caldeira-Leggett model.
Namely
\begin{equation}
\partial_t H_I = \left(-\frac{\dot{C}_q(t)}{C_q(t)}Q\right)\left(-\partial_Q H_I\right)
\end{equation}
which when $C_q\propto \sqrt{\eta(t)}$ leads to
\begin{eqnarray}
\delta W_{if} = \int \partial_t H_I dt &=& \int \left(-\frac{\dot{\eta}}{2\eta}Q\right)\big(
-\eta \dot{Q} -\frac{\dot{\eta}}{2\eta}Q 
\nonumber \\
&& +\sqrt{\frac{2\eta}{\beta}}\xi\big) dt
\quad\hbox{(from \cite{Aurell2017})}
\label{eq:W-if}
\end{eqnarray}
Summarizing, the effective motion of the system in the Caldeira-Leggett
model with time-dependent friction is 
\begin{equation}
\dot{Q}=\frac{P}{M}\quad \dot{P}=-\partial_Q V + F_S
\end{equation}
where the generalized Sekimoto force $F_S$ is
\begin{equation}
F_S = -\eta \dot{Q} -\frac{\dot{\eta}}{2\eta}Q +\sqrt{\frac{2\eta}{\beta}}\xi
\end{equation}
The change of internal energy is for this model
\begin{eqnarray}
\Delta U &=& \Delta H_S = \int \left(\partial_t H_S +\dot{P}\partial_P H_S + \dot{Q}\partial_Q H_S\right) dt \nonumber \\
         &=& \delta W^{(J)} + \int \frac{P}{M} F_S dt
\end{eqnarray}
and the work $\delta W_{if}$ from (\ref{eq:W-if}) is 
\begin{eqnarray}
\delta W_{if} &=& \int \left(-\frac{\dot{\eta}}{2\eta}Q\right) F_S dt
\label{eq:W-if-2}
\end{eqnarray}
Finally the heat is
\begin{eqnarray}
\delta Q &=& \delta H_B = \delta W - \Delta U = \nonumber \\
         &=& \int F_S \left(-\frac{\dot{\eta}}{2\eta}Q - \frac{P}{M}\right)  dt
\label{eq:W-if-2}
\end{eqnarray}
Work, heat and internal energy change are hence for this model in equal measure
functionals of the system history only.

The above approach can be generalized to interactions of the type 
\begin{equation}
H_I(Q,q,t) = A(Q)B(q)C(t)
\end{equation}
where $A(Q)$ is a known function of the system, and $C(t)$ is a known function of time.
The bath will then exert a force on the system as
\begin{equation}
-\partial_Q H_I(Q,q,t) = -\partial_Q A \left(B(q)C(t)\right)
\end{equation}
When the acceleration of the system can be measured, this force is a system observable since
\begin{equation}
-\partial_Q H_I(Q,q,t) = \dot{P}+\partial_Q H_S(Q,P,t)
\end{equation}
On the other hand
\begin{eqnarray}
\partial_t H_I(Q,q,t) &=& \left(A(Q)B(q)\right)\partial_t C \nonumber \\
&=& -\frac{\partial_t \log C}
{\partial_Q \log A} \left(-\partial_Q H_I(Q,q,t)\right)
\end{eqnarray}
The second contribution to the work is then a functional of system history as
\begin{eqnarray}
\int \partial_t H_I(Q,q,t) dt &=& \int \left(-\frac{\partial_t \log C}{\partial_Q \log A}\right) \nonumber \\
&& \left(dP+\partial_Q H_S(Q,P,t)dt\right)
\end{eqnarray}

\section{Strong-coupling system entropy, internal energy and free energy}
\label{app:Seifert-Anders}
This appendix contains the details of the transition from (\ref{eq:entropy-1}) to (\ref{eq:entropy-2})
in Section~\ref{sec:conditional} above.
We repeat the starting point as
\begin{eqnarray}
\Delta S_{TOT} &=& \Delta \left(-\log \rho_S\right) - \Delta\left(\log \sigma(y | x)\right)
\label{eq:entropy-1-app}
\end{eqnarray}
Using the assumption stated in~\cite{MillerAnders2017} below Eq. 17,
the two parts of the last term in (\ref{eq:entropy-1-app})
can be written
\begin{eqnarray}
\label{eq:log-term-1}
\log \sigma(y^i | x^i) &=& - \beta H_{TOT}^i+\beta H_{S}^i - \log\left<e^{-\beta H_I}\right>^i_B  \nonumber \\
\label{eq:log-term-2}
\log \sigma(y^f | x^f) &=& - \beta H_{TOT}^f+\beta H_{S}^f - \log\left<e^{-\beta H_I}\right>^f_B  \nonumber
\end{eqnarray}
where we have introduced the notation of \cite{Seifert2016}
\begin{equation}
\left<\cdots \right>_B = e^{\beta F_B} \int dy' e^{-\beta H_B(y')}\left(\cdots\right)
\end{equation}
We thus have a contribution to (\ref{eq:entropy-1-app}) as
\begin{eqnarray}
\label{eq:log-terms}
-\Delta \log \sigma(y | x) &=&  \beta \Delta H_{TOT} - \beta \Delta H_{S} \nonumber \\
 &&  +\Delta \log\left<e^{-\beta H_I}\right>_B  
\end{eqnarray}
The contributions of the free energy of the bath ($F_B$) cancel and do not contribute to 
(\ref{eq:log-terms}).

The difference $\Delta H_{TOT}$ in (\ref{eq:log-terms}) 
is the work, $\delta W$. Under the assumption that only $H_S$ depends explicitly on time
$\delta W$ is the Jarzynski work $\delta W^{(J)}$. The second difference $\Delta H_S$ in (\ref{eq:log-terms}) is the change of system internal energy
as usually defined. For many models of system-bath interaction that can also be
taken a functional of system history only. 

The logarithmic terms in (\ref{eq:log-terms}) 
can on the other hand be re-written 
\begin{eqnarray}
\log\left<e^{-\beta H_I}\right>_B &=& \beta^2\partial_{\beta}\left(-\frac{1}{\beta}\log\left<e^{-\beta H_I}\right>_B\right) \nonumber \\
&& +\beta\partial_{\beta} \log\left<e^{-\beta H_I}\right>_B 
\end{eqnarray}
The first term can be included in a strong-coupling system entropy
\begin{eqnarray}
\label{eq:S-Seifert}
\tilde{s}_S &=& -\log\rho_S +\beta^2\partial_{\beta}\left(-\frac{1}{\beta}\log\left<e^{-\beta H_I}\right>_B\right)  
\end{eqnarray}
while the second can be combined with the bare change of system internal energy as
\begin{eqnarray}
\tilde{u}_S &=& H_S -\partial_{\beta} \log\left<e^{-\beta H_I}\right>_B \nonumber \\
\label{eq:U-Seifert}
&=& \partial_{\beta}\left(\beta(H_S -\frac{1}{\beta}\log\left<e^{-\beta H_I}\right>_B\right)
\end{eqnarray}
With these definitions we hence have (\ref{eq:entropy-2}) which we copy also here as 
\begin{eqnarray}
\label{eq:entropy-2-app}
\Delta S_{TOT} &=& \Delta\tilde{s}_S + \beta\delta W - \beta \Delta \tilde{u}_S
\end{eqnarray}

The definitions of $\tilde{s}_S$ and ${u}_S$ can be related to a strong-coupling system free energy
\begin{equation}
\label{eq:F-Seifert}
\tilde{f}_S = \tilde{u}_S -\frac{1}{\beta} \tilde{s}_S = H_S + \frac{1}{\beta}\log\rho_S 
- \frac{1}{\beta} \log\left<e^{-\beta H_I}\right>_B
\end{equation}
through the standard thermodynamic relations (Legendre transforms)
\begin{eqnarray}
\tilde{u}_S &=& \partial_{\beta}\left(\beta\tilde{f}_S\right)= \tilde{f}_S +\beta\partial_{\beta}\tilde{f}_S\\
\tilde{s}_S &=&  \beta\left(\tilde{u}_S - \tilde{f}_S\right) = \beta\partial_{\beta}\left(\beta\tilde{f}_S\right) - \beta \tilde{f}_S 
\end{eqnarray}

\section{Details on time reversals in stochastic dynamics}
\label{app:stochastic-details}
This Appendix provides technical details for Section~\ref{sec:stochastic}
in the main text.
Kramers-Langevin equation 
$\dot{x}=\frac{p}{m}$ and $\dot{p}=-\partial_x V(x,t) - \gamma\frac{p}{m} + \sqrt{2k_B T\gamma}\xi$
is to be interpreted in the Stratonovich convention~\cite{ChG07}.
Over a small time interval $t$ to $t'=t+\epsilon$ this means
\begin{eqnarray}
x'-x &=& \epsilon \frac{\overline{p}}{m} \\
p'-p &=& -\epsilon\gamma \frac{\overline{p}}{m} -\epsilon \partial_x V(x,t) + \sqrt{2k_B T\gamma} \Delta \Xi
\end{eqnarray}
where $\overline{p}=\frac{p+p'}{2}$ and $\Delta \Xi$ is a centered normal variable of variance $\epsilon$.
Terms higher than $\epsilon$ have been suppressed.
It follows that the probability distribution of $p'$ conditioned on $p$ is  
\begin{widetext}
\begin{equation}
P(p'|p) = \frac{1}{(4\pi k_B T\gamma \epsilon)^{\frac{d}{2}}}
\exp\left(-\frac{\left( p'-p +\epsilon\gamma \frac{\overline{p}}{m} +\epsilon\partial_x V(x,t)\right)^2}{4 k_B T\gamma \epsilon}\right) 
 \left(1+\epsilon\frac{\gamma d}{2m}\right)
\end{equation}
\end{widetext}
where $d$ is dimension of space, and the last term arises from the Jacobian when transforming from $\Delta\Xi$ to $p'$.
\textit{Natural time reversal} of the Kramers-Langevin equation means
$\frac{dx^*}{dt^*}=\frac{p^*}{m}$ and $\frac{dp^*}{dt^*}=-\partial_{x^*} V(x^*,t^*) + \gamma\frac{p^*}{m} + \sqrt{2k_B T\gamma}\xi^*$
where $x^*_{t^*}=x_t$,  $p^*_{t^*}=-p_t$ and $\xi^*$ is a noise with the same characteristics as $\xi$.
The probability distribution of of $(p^*)'$ conditional on $p^*$ over a short time $t^*$ to  $(t^*)'=t^*+\epsilon$
is thus
\begin{widetext}
\begin{equation}
P((p^*)'|p^*) = \frac{1}{(4\pi k_B T\gamma \epsilon)^{\frac{d}{2}}}
\exp\left(-\frac{\left( (p^*)'-p^* -\epsilon\gamma \frac{\overline{p^*}}{m} +\epsilon\partial_{x^*} V(x^*,t^*)\right)^2}{4 k_B T\gamma \epsilon}\right) \left(1-\epsilon\frac{\gamma d}{2m}\right)
\end{equation}
\end{widetext} 
Inserting $(p^*)'=-p$ and $(p^*)=-p'$ one can form the ratio
\begin{equation}
\frac{P(p'|p)}{P((p^*)'|p^*)} = \left(1+\epsilon\frac{\gamma d}{m}\right) +{\cal O}(\epsilon^2)
\end{equation} 
which leads to an entropy production in the environment, over the whole process, as
\begin{equation}
\delta S_{env}^{natural} = \log \frac{P^F}{P^B} = (t_f - t_i) \frac{\gamma d}{m}
\label{eq:App-CG-formula-natural}
\end{equation} 
\textit{General time reversal} of the Kramers-Langevin equation as discussed in the main text means 
$\frac{dx^*}{dt^*}=\frac{p^*}{m}$ and $\frac{dp^*}{dt^*}= - \partial_{x^*} V + 2\partial_{x^*} V^{-} - \gamma\frac{p^*}{m} + \sqrt{2k_B T\gamma}\xi^{''}$
where $\xi^{''}$ is as above a noise with the same characteristics as $\xi$.
In this case the ratio of the two propagators over a short time interval is
\begin{widetext}
\begin{equation}
\frac{P(p'|p)}{P((p^*)'|p^*)} = \exp\left(\frac{1}{k_B T\gamma}\left(p'-p +\partial_x V^+\right)\left(-\gamma p/m -\partial_x V^-\right)\right)
\end{equation} 
\end{widetext} 
Introducing 
the notation of \cite{ChG07} that $u_+= -\gamma p/m -\partial_x V^-$
is the part of the drift field that transforms as a vector
and $u_-= -\partial_x V^+$ is the part that transforms as a pseudo-vector,
and identifying $D=k_B T\gamma$ as the diffusion coefficient, one has
\begin{eqnarray}
\delta S_{env}^{general} &=& \int \left(\dot{p}-u_-\right)\frac{1}{D} u_+ dt
\label{eq:App-CG-formula}
\end{eqnarray}
which is the formula quoted as (\ref{eq:CG-formula})
in the main text.
For \textit{canonical time reversal}, the special case of above when $V^-=0$,
a more detailed discussion along the same lines
as above can be found in~\cite{Celani12}.
Mathematically rigorous derivations of (\ref{eq:App-CG-formula-natural}) and 
(\ref{eq:App-CG-formula}), as well as other time reversals of diffusion
processes, can be found in~\cite{ChG07}.

\subsection{Microscopic model}
I will now show that (\ref{eq:App-CG-formula})
can also be derived as
the change of bath energy in an explicit model
of a bath as harmonic oscillators initially in thermal equilibrium.
The oscillators are labeled by their frequencies $\omega$,
have mass $m_{\omega}$ and density of states $f(\omega)$,
and interact with the system with coupling strength
$C_{\omega}$. 
An Ohmic spectrum that satisfies 
\begin{equation}
\frac{f(\omega) C^2(\omega)}{m_{\omega}} = \frac{2}{\pi}\gamma \omega^2
\label{eq:Ohmic}
\end{equation}
leads to Kramers-Langevin dynamics for the system with friction coefficient $\gamma$~\cite{Zwanzig,Bez1980,Hanggi1997}.

It is convenient to introduce terms for mappings:
\begin{description}
\item[${\cal I}$] is as before the mapping $(x,y,H)\to (x^*,y^*,H^*)$. On the system 
${\cal I}$ acts as in general time reversal above; on the level of the bath
the action of ${\cal I}$ is to be determined.
\item[${\cal T}$] is the forward evolution of the system and the bath from time $t_i$
and initial conditions $(x^i,y^i)$
to time $t_f$ and final conditions $(x^f,y^f)$ under Hamiltonian $H$.
\item[${\cal T}^*$] is the time reversed evolution of the system and the bath from time $t_i^*=0$
and initial conditions $((x^*)^i,(y^*)^i)$
to time $t_f^*=t_f-t_i$ and final conditions $((x^*)^f,(y^*)^f)$ under Hamiltonian $H^*$.
\item[${\cal F}$] is the determination of $(x^i,y^i)$, the initial conditions in the forward process,
in terms of $\{x_k\}_{k=0}^n$, the forward trajectory of the system.
Note that $x^i=x_0$ \textit{i.e.} this mapping is trivial on the system.
\item[${\cal F}^*$] is the determination of $(y^*)^i$, the initial conditions in the time-reversed process,
in terms of $\{x^*_k\}_{k=0}^n$, the time-reversed trajectory of the system. Also here $(x^*)^i=x^*_0$.
\end{description}

All mappings are assumed to be smooth and invertible as needed.
We can then define
\begin{eqnarray}
\label{eq:App-I-def-1}
{\cal I}(x^f,y^f) &=& {\cal F^*} {\cal I} {\cal F}^{-1} {\cal T}^{-1}(x^f,y^f)) \\
\label{eq:App-I-def-2}
{\cal I}((x^*)^f,(y^*)^f) &=& {\cal F} {\cal I} {\cal F^*}^{-1} {\cal T^*}^{-1}((x^*)^f,(y^*)^f)
\end{eqnarray}
In words the above says that 
the time-reversed final conditions of the bath,
in either process, are what they have to be as initial conditions
so that the whole trajectory of the system is time-reversed.
With these (formal) definitions ${\cal I}$ is an involution as illustrated 
by the following diagram:

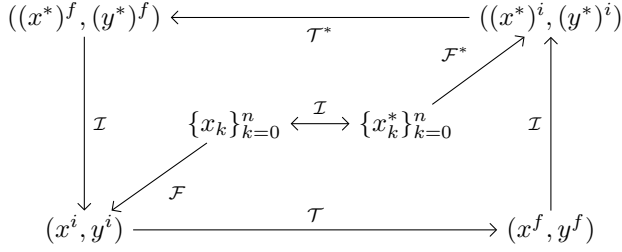
\begin{tikzpicture}
\matrix(m)[matrix of math nodes,
row sep=2.6em, column sep=0.3em,
text height=1.5ex, text depth=0.25ex]
{((x^*)^f,(y^*)^f) &                 &\quad &                  &  ((x^*)^i,(y^*)^i)  \\
                   & \{x_k\}_{k=0}^n &\quad &\{x^*_k\}_{k=0}^n &    \\
 (x^i,y^i)         &                 &\quad &                  & (x^f,y^f)\\};
\path[->,font=\scriptsize,>=angle 90]
(m-1-5) edge node[auto] {${\cal T}^*$} (m-1-1)
(m-3-1) edge node[auto] {${\cal T}$} (m-3-5)
(m-3-5) edge node[auto] {${\cal I}$} (m-1-5)
(m-1-1) edge node[auto] {${\cal I}$} (m-3-1);
\path[<->,font=\scriptsize,>=angle 90]
(m-2-2) edge node[auto] {${\cal I}$} (m-2-4);
\path[->,font=\scriptsize,>=angle 90]
(m-2-2) edge node[auto] {${\cal F}$} (m-3-1)
(m-2-4) edge node[auto] {${\cal F}^*$} (m-1-5);
\end{tikzpicture}

\subsection{Phase space volume}
To show that ${\cal I}$ preserves phase space volume
we have to consider the Jacobians corresponding to (\ref{eq:App-I-def-1})
and (\ref{eq:App-I-def-2}). To avoid under-counting in the continuously sampled
limit take the forward system path $\{x_k\}_{k=0}^n$
to be specified by initial system coordinates and momenta $x_0=(X^i,P^i)$
and $2n$ momenta increments $x_k=(\Delta P_{2k-1},\Delta P_{2k})$, and similarly 
for the time-reversed path.

The initial conditions of the bath are only reflected in the
noise term of the Kramers-Langevin equation; that is
\begin{eqnarray}
{\cal F}^{-1}&:&\sqrt{2k_B T\gamma}\xi = \int_0^{\infty} f(\omega) C(\omega) [q_{\omega} \cos\omega t + \nonumber \\
  && \qquad \frac{p_{\omega}}{m_{\omega}\omega}\sin\omega t ] d\omega
\label{eq:Zwanzig-noise}
\end{eqnarray}
and similarly for the backward process
\begin{eqnarray}
{\cal F^*}^{-1}&:& \sqrt{2k_B T\gamma}\xi^{''} = \int_0^{\infty} f(\omega) C(\omega) [q^*_{\omega} \cos\omega t^* + \nonumber \\
  &&\qquad  \frac{p^*_{\omega}}{m_{\omega}\omega}\sin\omega t^* ] d\omega
\label{eq:Zwanzig-noise-bw}
\end{eqnarray}
Eq.~(\ref{eq:Zwanzig-noise}) determines how the momentum increments ($\Delta P_{k}, k>0$)
depend on the initial conditions of the bath ($q_{\omega}, p_{\omega}$), and analogously for the time-reversed path.
The initial conditions of the paths can be solved for
by inverse Fourier transform 
\begin{equation}
{\cal F} : \left\{    \begin{array}{lcl} q_{\omega} &=& \frac{1}{\pi}  \frac{1}{f(\omega) C(\omega)} \int (\dot{p} +\partial_x V +\gamma p/m) \cos\omega t d t \\  
                              p_{\omega} &=& \frac{1}{\pi}  \frac{m_{\omega}\omega}{f(\omega) C(\omega)}\int (\dot{p} +\partial_x V +\gamma p/m) \sin\omega t d t 
                \end{array} \right.
\label{eq:inv-Zwanzig-noise-1}
\end{equation}
and similarly
\begin{equation}
{\cal F^*} :  \left\{ \begin{array}{lcl} q^*_{\omega} &=& \frac{1}{\pi}  \frac{1}{f(\omega) C(\omega)} \int (\dot{p} +\partial_x V -2\partial_x V^-  \\
                                 &&\qquad -\gamma p/m) \cos\omega t^* d t^* \\
                               p^*_{\omega} &=& \frac{1}{\pi}  \frac{m_{\omega}\omega}{f(\omega) C(\omega)} \int (\dot{p} +\partial_x V -2\partial_x V^- \\
                                 &&\qquad -\gamma p/m) \sin\omega t^* d t^*
            \end{array}\right. 
\label{eq:inv-Zwanzig-noise-2}
\end{equation}
Eq.~(\ref{eq:App-I-def-1}) defines the determinant of the Jacobian of ${\cal I}$ as
\begin{eqnarray}
\label{eq:App-I-det}
|\frac{\partial {\cal I}(x^f,y^f)}{\partial (x^f,y^f)}|  &=& |\frac{\partial ((x^*)^i,(y^*)^i)}{\partial \{x^*_k\}_{k=0}^n}|\,\cdot\,  |\frac{\partial \{x^*_k\}_{k=0}^n}{\partial \{x_k\}_{k=0}^n}|\,\cdot 
\nonumber \\  
&&\quad \cdot\,|\frac{\partial \{x_k\}_{k=0}^n}{\partial (x^i,y^i)}|\,\cdot\, |\frac{\partial (x^i,y^i)}{\partial (x^f,y^f)}| \nonumber \\  
&=& |\frac{\partial \{x^*_k\}_{k=1}^n}{\partial (y^*)^i}|^{-1}\,\cdot\, |\frac{\partial \{x_k\}_{k=1}^n}{\partial y^i}|
\end{eqnarray}
In above has been used that $|\frac{\partial \{x^*_k\}_{k=0}^n}{\partial \{x_k\}_{k=0}^n}|$
is one because ${\cal I}$ preserves system volume,
that $\frac{\partial (x^i,y^i)}{\partial (x^f,y^f)}$ is one because the 
full dynamics is conservative, and that ${\cal F}$ acts trivially on the system.
The whole expression is finally one because by
(\ref{eq:Zwanzig-noise}) and (\ref{eq:Zwanzig-noise-bw})
the two Jacobians 
$\frac{\partial \{x^*_k\}_{k=1}^n}{\partial (y^*)^i}$ and $\frac{\partial \{x_k\}_{k=1}^n}{\partial y^i}$
are the same.

\subsection{Change of bath energy}
Finally we consider the energy change of the bath between the starting positions of the backward and forward process:
\begin{eqnarray}
\Delta H_B &=& \int_0^{\infty} f(\omega) ( \frac{1}{2m_{\omega}}((p^*_{\omega})^2-(p_{\omega})^2 +\nonumber \\
&&\quad \frac{1}{2}m_{\omega}\omega^2 ((q^*_{\omega})^2-(q_{\omega})^2)) d\omega
\label{eq:energy-change}
\end{eqnarray}
where the contributions from a given $\omega$ are
\begin{eqnarray}
&&\frac{(p^*_{\omega})^2-p^2_{\omega}}{2m_{\omega}}+ \frac{1}{2} m_{\omega}\omega^2  ((q^*_{\omega})^2 -q^2_{\omega} ) =\nonumber \\
&=& \left(\frac{m_{\omega}\omega}{\pi f(\omega)C(\omega)}\right)^2 
\int\int \cos\omega (t-t')[-2(\dot{p}+\partial_x V^+) \cdot \nonumber \\
&&\quad (\gamma p/m +\partial_x V^-)^{'} - 2(\dot{p}+\partial_x V^+)^{'} \cdot \nonumber \\
&&\qquad (\gamma p/m +\partial_x V^-)]\,dt \, dt'
\label{eq:bath-change-2}
\end{eqnarray} 
In above primed quantities refer to to time $t'$ and unprimed to time $t$. 
Using (\ref{eq:Ohmic}), the notation in (\ref{eq:App-CG-formula}), (\ref{eq:energy-change})
and $\int \cos\omega (t-t') d\omega = 2\pi \delta(t-t')$
this leads to
\begin{equation}
\Delta H_B = \int \left(\dot{p}-u_-\right)\frac{1}{\gamma} u_+ dt
\label{eq:App-CG-formula-heat}
\end{equation}
which is Eq.~\ref{eq:CG-formula-heat} in main text.

\bibliography{fluctuations}%
\end{document}